\documentclass[12pt]{article}

\usepackage{epsfig,amssymb,euscript}
\usepackage{amsmath,amscd}

\numberwithin{equation}{section}
\newcommand{\be}{\begin{eqnarray}}
\newcommand{\ee}{\end{eqnarray}}
\newcommand{\bea}{\begin{eqnarray}}
\newcommand{\eea}{\end{eqnarray}}
\newcommand{\ba}{\begin{array}}
\newcommand{\ea}{\end{array}}
\newcommand{\nn}{\nonumber \\}

\begin{document}

\begin{titlepage}

\vfill


\vfill

\begin{center}

\baselineskip=16pt

{\Large\bf Isometries of Half Supersymmetric Time-Like Solutions In Five Dimensions}

\vskip 1.3cm

Jan B. Gutowski$^1$ and Wafic A. Sabra $^2$

\vskip 1cm

{\small{\it
$^1$DAMTP, Centre for Mathematical Sciences\\
University of Cambridge\\
Wilberforce Road, Cambridge, CB3 0WA, UK\\
Email: J.B.Gutowski@damtp.cam.ac.uk}}

\vskip .6cm {\small{\it
$^2$ Centre for Advanced Mathematical Sciences and Physics Department\\
American University of Beirut\\ Lebanon \\
Email: ws00@aub.edu.lb \\}}

\end{center}

\vfill

\begin{center}
\textbf{Abstract}
\end{center}

\begin{quote}
Spinorial geometry techniques have recently been 
used to classify all half supersymmetric solutions in gauged five dimensional
supergravity with vector multiplets. In this paper we
consider solutions for which at least one of the Killing 
spinors generates a timelike Killing vector.
We obtain coordinate transformations 
which considerably simplify the solutions, and in a number of cases, we 
obtain explicitly some additional 
Killing vectors which were hidden in the original analysis.
\end{quote}
\begin{quote}
\end{quote}
\vfill

\end{titlepage}

\section{Introduction}

The classification of supersymmetric supergravity solutions is
of interest in a number of contexts, such as the
conjectured AdS/CFT correspondence, the microscopic analysis of black
hole entropy, and a deeper understanding of string theory dualities.
In particular, in five dimensions, considerable progress has been made in the classification and
the analysis of solutions preserving various fractions of supersymmetry of 
$N=2$ gauged supergravity \cite{electric}-\cite{harimulti}. In these
theories, the only solution which preserves all of the supersymmetries is 
$AdS_{5}$ with vanishing gauge field strengths and constant scalars. The
Killing spinor equations when expressed in terms of Dirac spinors are linear
over $\mathbb{C}$, implying that supersymmetric solutions preserve $2$, $4,6$
or $8$ of the supersymmetries. However in the ungauged theory it was found
that supersymmetric solutions can only preserve $4$ or $8$ of
supersymmetries \cite{pakis, janunique}. The first supersymmetric solutions
which were constructed \cite{electric} have naked singularities or naked
closed time-like curves. Domain walls and magnetic strings were also
constructed in \cite{magnetic}. Later, a more systematic approach, motivated
by the results of  Tod \cite{tod}, was employed to construct and classify
supersymmetric solutions. Quarter supersymmetric solutions of minimal gauged 
$N=2,$ $D=5$ supergravity where classified in \cite{gaungut}. The first
asymptotically $AdS_{5}$ solutions with no closed time loops or naked
singularities were constructed for the minimal supergravity theory in 
\cite{gutowski1}. Generalisations to solutions with null Killing spinors as well
as to solutions with vector multiplets were later given in \cite{gutowski2,
gs}.

Half supersymmetric solutions posses two Dirac Killing spinors thus enabling
the construction of two Killing vectors as bilinears in the Killing spinors.
These vectors could be either time-like or null and thus there exists three
classes of solutions, depending on the nature of the Killing spinors and
vectors considered. All of these solutions have been fully classified 
in \cite{halfgs, halfjgs}, using the spinorial geometry method.
In this approach to the analysis of the Killing spinor equations,
the Killing spinors are expressed in terms of differential forms 
\cite{lawson, wang, harvey}, which are further simplified by
making appropriately chosen gauge transformations.
These techniques were originally developed to
analyse more complicated supergravity theories in
ten and eleven dimensions, see for example \cite{papadgran2005a,
papadgran2005b, papadgran2006a, papadgran2006b};
it is also straightforward to use them to analyse lower dimensional
theories as well. The remaining class of supersymmetric 
solutions, preserving 3/4 of the supersymmetry, has also been
considered recently in \cite{preons} where it
was shown, again using spinorial geometry techniques,
that these solutions are merely cosets of $AdS_{5}.$

It must be noted that supersymmetric solutions are not automatically
solutions of the equations of motion. For the $1/4$ supersymmetric time-like
solutions, one must solve the gauge equations and the Bianchi identities in
addition to the Killing spinor equations. In the null case, however, one
must additionally solve one of the components of the Einstein equations of
motion. For the $1/2$ supersymmetric solutions, considered in our present
work, it was found that supersymmetry together with Bianchi identities are
sufficient to imply that all the Einstein, gauge and scalar equations of
motion are all satisfied.

Our present work is concerned with half supersymmetric solutions of gauged 
$N=2,$ $D=5$ supergravity. We examine the time-like
solutions of \cite{halfgs}, and find coordinate transformations which simplify
these solutions and make it possible to extract some hidden symmetries. Our
work is organised as follows. In section two we present the basics of the
theories we are studying and their half supersymmetric solutions which are
presented in six classes. As three of these classes are already presented in 
\cite{halfgs} in the most explicit possible form, we focus on the other three classes
and find coordinate transformations that simplify them. This simplification
is presented in section 3.

\section{Half Supersymmetric Timelike Solutions}

In this section, we present a summary of all half supersymmetric solutions
of gauged $N=2$, $D=5$ supergravity, for which at least one of the Killing
spinors generates a timelike Killing vector. Such solutions were completely
classified in \cite{halfgs} and are specified in terms of a spacetime
metric, together with a number of scalars $X^{I}$ and 2-form gauge field
strengths $F^{I}$. In the classification of these solutions, the Killing
spinors were expressed as differential forms on $\Lambda ^{\ast }(\mathbb{R}
^{2})\otimes $ $\mathbb{C}$.  
One starts with the most generic form 
for a symplectic Majorana spinor, then, by applying 
appropriately chosen gauge transformations,
the spinor can be written in a number of particularly simple canonical forms.
The conditions for quarter supersymmetric solutions with a time-like Killing
vector (generated from the Killing spinor) are then derived. The conditions are then substituted into the
Killing spinor equations, acting on the second spinor,
and further constraints on the K\"ahler
base are then determined. Using the integrability conditions of the Killing
spinor equations, it was demonstrated that for a given background preserving
at least half of the supersymmetry, where at least one of the Killing
spinors generates a time-like Killing vector, all of the Einstein, gauge and
scalar field equations of motion hold automatically provided that the
Bianchi identity is satisfied. 

In listing all possible solutions, it will be particularly useful to
introduce co-ordinates $t,x$ and decompose the mostly negative signature
metric as 
\begin{equation}
ds^{2}=f^{4}(dt+\Omega )^{2}-f^{-2}ds_{\mathbf{B}}^{2}\text{ ,}
\end{equation}
where ${\frac{\partial }{\partial t}}$ is a Killing vector which is a
symmetry of the whole solution, $ds_{\mathbf{B}}^{2}$ is the metric on a
4-manifold $\mathbf{B}$ which is constrained by supersymmetry to be 
K\"ahler; $f$ is a $t$-independent function and $\Omega $ is a $t$-independent
1-form on $\mathbf{B}$. The bosonic action of the theory is 
\begin{equation}
S={\frac{1}{16\pi G}}\int \left( -{}R+2\chi ^{2}{\mathcal{V}}\right) {
\mathcal{\ast }}1+Q_{IJ}\left( dX^{I}\wedge \star dX^{J}-F^{I}\wedge \ast
F^{J}\right) -{\frac{C_{IJK}}{6}}F^{I}\wedge F^{J}\wedge A^{K}
\label{action}
\end{equation}
where $I,J,K$ take values $1,\ldots ,n$. The scalar fields $X^{I}$ are
subject to the constraint 
\begin{equation}
X_{I}X^{I}=1\ ,
\end{equation}
where
\be
X_I = {1 \over 6} C_{IJK} X^J X^K \ .
\ee
The coupling $Q_{IJ}$ and the scalar potential depend on the scalars via 
\begin{eqnarray}
Q_{IJ} &=&{\frac{9}{2}}X_{I}X_{J}-{\frac{1}{2}}C_{IJK}X^{K}, \\
{\mathcal{V}} &=&9V_{I}V_{J}(X^{I}X^{J}-{\frac{1}{2}}Q^{IJ})
\end{eqnarray}
where $V_{I}$ are constants.

Bosonic backgrounds are said to be supersymmetric if there exists a spinor $
\epsilon ^{a}$ for which the supersymmetry variations of the gravitino and
dilatino vanish in the given background. For the gravitino this requires

\begin{equation}
\left[ \nabla _{\mu }+\frac{1}{8}X_{I}\left( \gamma _{\mu }F^{I}{}_{\rho
\sigma }\gamma ^{\rho \sigma }-{6}F^{I}{}_{\mu \rho }\gamma ^{\rho }\right) 
\right] \epsilon ^{a}-{\frac{\chi }{2}}V_{I}(X^{I} \gamma
_{\mu}-3A^{I}{}_{\mu })\epsilon ^{ab}\epsilon ^{b}=0,  \label{eqn:grav}
\end{equation}
and for the dilatino it requires

\begin{equation}
\left[ {\frac{1}{4}}\left( Q_{IJ}\gamma ^{\mu \nu }F^{J}{}_{\mu \nu }+{3}
\gamma ^{\mu }\nabla _{\mu }X_{I}\right) \epsilon ^{a}-{\frac{3\chi }{2}}
V_{I}\epsilon ^{ab}\epsilon ^{b}\right] {\partial _{r}X^{I}}=0\,.
\label{eqn:dil}
\end{equation}
The Einstein equations derived from (\ref{action}) are given by

\begin{equation}
{}R_{\alpha \beta }+Q_{IJ}\left( F^{I}{}_{\alpha \lambda }F^{J}{}_{\beta
}{}^{\lambda }-\nabla _{\alpha }X^{I}\nabla _{\beta }X^{J}-{\frac{1}{6}}
g_{\alpha \beta }F^{I}{}_{\mu \nu }F^{J\mu \nu }\right) -{\frac{2}{3}}
g_{\alpha \beta }\chi ^{2}\mathcal{V}=0\,.  \label{eqn:ein}
\end{equation}
The Maxwell equations are

\begin{equation}
d\left( Q_{IJ}\star F^{J}\right) =-{\frac{1}{4}}C_{IJK}F^{J}\wedge F^{K}\,.
\label{eqn:gauge}
\end{equation}
The scalar equations are

\begin{eqnarray}
&&\bigg[{-}d(\star dX_{I})+\left( X_{M}X^{P}C_{NPI}-{\frac{1}{6}}
C_{MNI}\right) (F^{M}\wedge \star F^{N}-dX^{M}\wedge \star dX^{N})  \notag \\
&&-{\frac{3}{2}}\chi ^{2}V_{M}V_{N}Q^{ML}Q^{NP}C_{LPI}\mathrm{dvol}\bigg]{
\partial _{r}X^{I}}=0\,.\qquad 
\end{eqnarray}

We refer the reader to \cite{halfgs} for the details of the analysis of the Killing spinor equations.
The solutions can be divided into six classes:

\begin{itemize}
\item[(1)] The K\"ahler base metric has co-ordinates $\tau, \eta, u ,v$ and
is given by

\begin{eqnarray}
ds_{\mathbf{B}}^{2} &=&H^{2}\left( d\tau +\eta \left( {\frac{\partial Y}{
\partial u}}+H^{2}\sin Y\right) du+\eta {\frac{\partial Y}{\partial v}}
dv\right) ^{2}+{\frac{dv^{2}}{H^{2}}}+H^{2}v^{2}\sin ^{2}Ydu^{2}  \notag \\
&+&H^{2}\left( d\eta -\eta {\cot Y}{\frac{\partial Y}{\partial u}}du-\eta
\left( {\cot Y}{\frac{\partial Y}{\partial v}}+{\frac{1}{v}}\right)
dv\right) ^{2}
\end{eqnarray}
and

\begin{eqnarray}
\Omega &=&-{\frac{1}{2cv}}(H^{2}+\frac{c^{2}v^{2}}{f^{6}})d\tau +{\frac{\eta 
}{cv}}({\frac{1}{2}}\theta \sin Y-H^{2}{\frac{\partial Y}{\partial v}})dv 
\notag \\
&&-\eta \left( {\frac{H^{2}}{cv}}({\frac{\partial Y}{\partial u}}+H^{2}\sin
Y)+{\frac{1}{2}}\sin Y(\frac{{\theta }H^{2}}{c}\cos Y+\frac{c}{f^{6}}H^{2}v-{
\frac{1}{cv}}H^{4})\right) du
\nn
\end{eqnarray}

and the scalars $X^{I}$ and function $f$ are constrained by

\begin{eqnarray}  \label{scalars1}
\frac{\partial }{\partial u}(\frac{X_{I}}{f^{2}}) &=& \frac{{\chi }
H^{2}V_{I}\sin ^{2}Y}{c},  \notag \\
{\frac{\partial }{\partial v}}(\frac{X_{I}}{f^{2}}) &=&\frac{1}{v}\left( 
\frac{{\chi }V_{I}(\cos Y-1)}{c}-\frac{X_{I}}{f^{2}}\right) .
\end{eqnarray}

For this solution, the scalars $X^I$ and $f$ depend only on $u, v$; $c,
\theta$ are constants with $c \neq 0$, and $H$, $Y$ are functions of $u, v$
such that $\sin Y \neq 0$, which are constrained by

\begin{eqnarray}
{\frac{\partial H^{2}}{\partial u}} &=&H^{2}v\sin ^{2}Y\left( 3\frac{\chi
cvV_{I}X^{I}}{f^{4}}-\theta \right) ,  \notag \\
{\frac{\partial H^{2}}{\partial v}} &=&-\frac{cv}{f^{4}}\left( 3\chi
V_{I}X^{I}+\frac{c}{f^{2}}\right) +\cos Y(3\frac{\chi cvV_{I}X^{I}}{f^{4}}
-\theta ),  \label{Hcons2}
\end{eqnarray}
and

\begin{eqnarray}
{\frac{\partial Y}{\partial u}} &=&\sin Y\left( -H^{2}+3\frac{\chi
cv^{2}V_{I}X^{I}}{f^{4}}+\frac{c^{2}v^{2}}{f^{6}}\right) +\frac{v}{2}\sin
2Y\left( 3\frac{\chi cvV_{I}X^{I}}{f^{4}}-\theta \right) ,  \notag \\
{\frac{\partial Y}{\partial v}} &=&-\frac{1}{H^{2}}\sin Y\left( 3\frac{\chi
cvV_{I}X^{I}}{f^{4}}-\theta \right) .  \label{Ycons2}
\end{eqnarray}

If $\theta \neq 0$, then one can integrate up the constraints ({\ref
{scalars1}}) to obtain

\begin{equation}
X_{I}=f^{2}\left( {\frac{q_{I}}{v}+\frac{\chi }{c}}\left( \frac{c^{2}v}{
f^{6}\theta }-{\frac{H^{2}}{\theta v}}-1\right) V_{I}\right) .
\end{equation}

The gauge field strengths are given by 
\begin{eqnarray}
F^{I} &=&d\big[f^{2}X^{I}(dt+\Omega )+\frac{cvX^{I}}{f^{4}}\left( d\tau
+\eta H^{2}\sin Ydu\right)   \notag \\
&&-\frac{3\chi \eta }{f^{2}}\sin Y(X^{I}X^{J}-{\frac{1}{2}}
Q^{IJ})V_{J}(-H^{2}v(1+\cos Y)du+dv)\big].
\end{eqnarray}

\item[(2)] For the second class of solutions, one can choose a co-ordinate $v
$ on $\mathbf{B}$ together with three $v$-independent 1-forms $\sigma ^{i}$ (
$i=1,2,3$) on $\mathbf{B}$ orthogonal to ${\frac{\partial }{\partial v}}$.
One also has constants $c$, $\theta $ ($c\neq 0$) and the solution takes one
of three types according as $c\theta $ is negative, zero or positive. If $
\theta \neq 0$ then 
\begin{equation}
ds_{\mathbf{B}}^{2}={\frac{1}{\theta v+c^{2}v^{2}f^{-6}}}dv^{2}+{\frac{v}{
\theta ^{2}}}(\theta +c^{2}vf^{-6})(\sigma ^{1})^{2}+{\frac{v}{|\theta |}}
((\sigma ^{2})^{2}+(\sigma ^{3})^{2}),
\end{equation}
and if $\theta =0$, 
\begin{equation}
ds_{\mathbf{B}}^{2}={\frac{1}{c^{2}v^{2}f^{-6}}}dv^{2}+4c^{8}f^{-6}v^{2}(
\sigma ^{1})^{2}+2c^{3}v((\sigma ^{2})^{2}+(\sigma ^{3})^{2}).
\end{equation}

The 1-forms $\sigma ^{i}$ satisfy

\begin{tabular}{lr}
$d\sigma ^{i}=-{\frac{1}{2}}\epsilon _{ijk}\sigma ^{j}\wedge \sigma ^{k}$ & 
: \qquad\ \ \textrm{if } $c\theta >0,$ \\ 
$d\sigma ^{1}=\sigma ^{2}\wedge \sigma ^{3}$, $d\sigma ^{2}=\sigma
^{1}\wedge \sigma ^{3}$, $d\sigma ^{3}=-\sigma ^{1}\wedge \sigma ^{2}$ & :
\qquad\ \ \textrm{if } $c\theta <0,$ \\ 
$d\sigma ^{1}=\sigma ^{2}\wedge \sigma ^{3}$, $d\sigma ^{2}=d\sigma ^{3}=0$
\ \  & : \qquad\ \ \textrm{if } $c\theta =0,$ \\ 
& 
\end{tabular}

If $\theta \neq 0$ then 
\begin{equation}
\Omega =-{\frac{cv}{\theta }}f^{-6}\sigma ^{1},
\end{equation}
whereas if $\theta =0$ then 
\begin{equation}
\Omega =2c^{4}vf^{-6}\sigma ^{1}.
\end{equation}

In all cases, the scalars $f$ and $X^{I}$ are constrained by 
\begin{equation}
X_{I}={\frac{f^{2}}{c}}(-2\chi V_{I}+{\frac{\rho _{I}}{\sqrt{2}v}})
\end{equation}
for constants $\rho _{I}$ and 
\begin{equation}
F^{I}=d(f^{2}X^{I}dt).
\end{equation}

\item[(3)] For the third class of solutions one can again choose a
co-ordinate $v$ on $\mathbf{B}$ together with three $v$-independent 1-forms $
\sigma ^{i}$ ($i=1,2,3$) on $\mathbf{B}$, orthogonal to ${\frac{\partial }{
\partial v}}$. For these solutions, the scalars $X^{I}$ are constant, and it
is convenient to define 
\begin{equation}
\Lambda =c\theta +9\sqrt{2}\chi ^{2}(X^{I}X^{J}-{\frac{1}{2}}
Q^{IJ})V_{I}V_{J}
\end{equation}
for constants $c$, $\theta $ ($c\neq 0$). The scalar $f$ is given by

\begin{equation}
f^{2}=\sqrt{2}cv.
\end{equation}

The solution takes one of three types. If $\Lambda \neq 0$ then 
\begin{eqnarray}
ds_{\mathbf{B}}^{2} &=&{\frac{1}{({\frac{1}{2\sqrt{2}cv}}-\theta v+{\frac{
3\chi }{c}}V_{I}X^{I})}}dv^{2}+{\frac{c^{2}}{\Lambda ^{2}}}({\frac{1}{2\sqrt{
2}cv}}-\theta v+{\frac{3\chi }{c}}V_{I}X^{I})(\sigma ^{1})^{2}  \notag \\
&+&{\frac{cv}{|\Lambda |}}((\sigma ^{2})^{2}+(\sigma ^{3})^{2}),
\end{eqnarray}
and if $\Lambda =0$, 
\begin{eqnarray}
ds_{\mathbf{B}}^{2} &=&{\frac{1}{({\frac{1}{2\sqrt{2}cv}}-\theta v+{\frac{
3\chi }{c}}V_{I}X^{I})}}dv^{2}+2c^{2}({\frac{1}{2\sqrt{2}cv}}-\theta v+{
\frac{3\chi }{c}}V_{I}X^{I})(\sigma ^{1})^{2}  \notag \\
&+&\sqrt{2}cv((\sigma ^{2})^{2}+(\sigma ^{3})^{2}),
\end{eqnarray}

The 1-forms $\sigma^i$ satisfy

\begin{tabular}{lr}
$d\sigma ^{i}=-{\frac{1}{2}}\epsilon _{ijk}\sigma ^{j}\wedge \sigma ^{k}$ & 
: \qquad\ \ \textrm{if } $\Lambda >0,$ \\ 
$d\sigma ^{1}=\sigma ^{2}\wedge \sigma ^{3}$, $d\sigma ^{2}=\sigma
^{1}\wedge \sigma ^{3}$, $d\sigma ^{3}=-\sigma ^{1}\wedge \sigma ^{2}$ & :
\qquad\ \ \textrm{if } $\Lambda <0,$ \\ 
$d\sigma ^{1}=\sigma ^{2}\wedge \sigma ^{3}$, $d\sigma ^{2}=d\sigma ^{3}=0$
\ \  & : \qquad\ \ \textrm{if } $\Lambda =0.$ \\ 
& 
\end{tabular}

If $\Lambda \neq 0$ then 
\begin{eqnarray}
\Omega &=&{\frac{1}{\Lambda cv^{2}}}({\frac{1}{2\sqrt{2}}}+{\frac{3\chi v}{2}
}V_{I}X^{I})\sigma ^{1},  \notag \\
F^{I} &=&d\big(\sqrt{2}cvX^{I}dt+{\frac{3\chi }{\sqrt{2}\Lambda }}
(Q^{IJ}-X^{I}X^{J})V_{J}\sigma ^{1}\big)
\end{eqnarray}
whereas if $\Lambda =0$, then 
\begin{eqnarray}
\Omega &=&{\frac{\sqrt{2}}{cv^{2}}}({\frac{1}{2\sqrt{2}}}+{\frac{3\chi v}{2}}
V_{I}X^{I})\sigma ^{1},  \notag \\
F^{I} &=&d\big(\sqrt{2}cvX^{I}dt+3\chi (Q^{IJ}-X^{I}X^{J})V_{J}\sigma ^{1}
\big).
\end{eqnarray}

\item[(4)] For the fourth class of solution, the scalars $X^{I}$ are
constant ($V_{I}X^{I}\neq 0$), and 
\begin{equation}
f=1.
\end{equation}

The K\"{a}hler base metric is the product of two 2-manifolds 
\begin{equation}
ds_{\mathbf{B}}^{2}=ds^{2}(M_{1})+ds^{2}(M_{2})
\end{equation}
where $M_{1}$ is $\mathbb{H}^{2}$ with Ricci scalar $R=-18\chi
^{2}(V_{I}X^{I})^{2}$, and $M_{2}$ is $\mathbb{H}^{2}$, $\mathbb{R}^{2}$ or $
\mathbb{S}^{2}$ with Ricci scalar $R=18\chi ^{2}(Q^{IJ}-X^{I}X^{J})V_{I}V_{J}
$. In addition, we have
\begin{equation}
d\Omega =3\chi V_{I}X^{I}\mathrm{dvol\ }(M_{1}),\qquad F^{I}=3\chi
(X^{I}X^{J}-Q^{IJ})V_{J}\mathrm{dvol\ }(M_{2})
\end{equation}

where $\mathrm{dvol\ }(M_{1})$, $\mathrm{dvol\ }(M_{2})$ are the volume
forms of $M_{1}$, $M_{2}$.

\item[(5)] For the fifth class of solution, one takes co-ordinates $\psi
,\phi ,x^{1},x^{2}$ on the base space $\mathbf{B}$, whose metric is given by 
\begin{equation}
ds_{\mathbf{B}}^{2}=e^{\sqrt{2}\varrho ^{2}\psi }\left[ (d\phi +\beta
)^{2}+d\psi ^{2}+T^{2}((dx^{1})^{2}+(dx^{2})^{2})\right] 
\end{equation}
where $\beta =\beta _{i}(x^{1},x^{2})dx^{i}$, $\varrho $ is a non-zero
constant and $T=T(x^{1},x^{2})$ is a scalar. $\beta $ is constrained by the
relation 
\begin{equation*}
T^{2}={\frac{1}{\sqrt{2}\varrho ^{2}}}\left( {\frac{\partial \beta _{2}}{
\partial x^{1}}}-{\frac{\partial \beta _{1}}{\partial x^{2}}}\right) .
\end{equation*}
\end{itemize}

The scalars $X^{I}$ depend only on $x^{1},x^{2}$, and 
\begin{equation}
f=e^{{\frac{1}{\sqrt{2}}}\varrho ^{2}\psi }u
\end{equation}
for a function $u$ which depends only on $x^{1},x^{2}$. There also exist two
purely imaginary functions $\mathcal{G}$, $\mathcal{H}$ which depend only on 
$x^{1},x^{2}$ and satisfy the constraints

\begin{equation}
{\frac{\partial }{\partial x^{2}}}(T\mathcal{H})={\frac{\partial }{\partial
x^{1}}}(T\mathcal{G}),
\end{equation}
and 
\begin{equation}
{\frac{\partial }{\partial x^{1}}}(\frac{\mathcal{H}}{T})={\frac{\partial }{
\partial x^{2}}}(\frac{\mathcal{G}}{T}),\qquad {\frac{\partial }{\partial
x^{1}}}(\frac{\mathcal{G}}{T})=-{\frac{\partial }{\partial x^{2}}}(\frac{
\mathcal{H}}{T}).
\end{equation}
The scalars $u$ and $X^{I}$ are constrained by

\begin{equation}
X_{I}=u^{2}q_{I}+{\chi u}^{2}\left( -{\frac{i}{\sqrt{2}\varrho ^{3}}}(\frac{1
}{T}{\frac{\partial \mathcal{G}}{\partial x^{2}}}+\frac{\mathcal{H}}{T^{2}}{
\frac{\partial T}{\partial x^{1}}})+\frac{3\chi }{\varrho ^{4}u^{2}}
V_{J}X^{J}\right) V_{I}
\end{equation}
for constant $q_{I}$, and $T$ satisfies the equation

\begin{equation}
\Box \log T+2\varrho ^{4}T^{2}=18\chi ^{2}(X^{I}X^{J}-{\frac{1}{2}}
Q^{IJ})V_{I}V_{J}\frac{T^{2}}{u^{2}}
\end{equation}
where $\Box =({\frac{\partial }{\partial x^{1}}})^{2}+({\frac{\partial }{
\partial x^{2}}})^{2}$ is the Laplacian on $\mathbb{R}^{2}$. Finally, $
\Omega $ is given by 
\begin{eqnarray}
\Omega &=&-{\frac{e^{-\sqrt{2}\varrho ^{2}\psi }}{\sqrt{2}}}\big[iT\varrho (-
\mathcal{G}dx^{1}+\mathcal{H}dx^{2})  \notag \\
&+&\big(-{\frac{i}{\sqrt{2}}}\frac{1}{\varrho }(\frac{1}{T}{\frac{\partial 
\mathcal{G}}{\partial x^{2}}}+\frac{\mathcal{H}}{T^{2}}{\frac{\partial T}{
\partial x^{1}}})+\frac{3\chi }{\varrho ^{2}u^{2}}V_{I}X^{I}\big)(d\phi
+\beta )\big]
\end{eqnarray}
and the gauge field strengths are 
\begin{equation}
F^{I}=d\left( f^{2}X^{I}(dt+\Omega )\right) +6\chi (X^{I}X^{J}-{\frac{1}{2}}
Q^{IJ})V_{J}\frac{T^{2}}{u^{2}}dx^{1}\wedge dx^{2}.
\end{equation}

\begin{itemize}
\item[(6)] For the sixth class of solutions, it is again convenient to
introduce coordinates $\phi $, $\psi $, $x^{1}$, $x^{2}$ on the K\"{a}hler
base. The base space metric is then 
\begin{equation*}
ds_{\mathbf{B}}^{2}=d\phi ^{2}+d\psi ^{2}+T^{2}((dx^{1})^{2}+(dx^{2})^{2})
\end{equation*}
where $T=T(x^{1},x^{2})$. The scalars $f$ and $X^{I}$ depend only on $
x^{1},x^{2}$. Again, there also exist two purely imaginary functions $
\mathcal{G}$, $\mathcal{H}$ which depend only on $x^{1},x^{2}$ and satisfy
the constraints
\end{itemize}

\begin{equation}
{\frac{\partial }{\partial x^{2}}}(T\mathcal{H})={\frac{\partial }{\partial
x^{1}}}(T\mathcal{G}),  \label{difconstr1}
\end{equation}
and 
\begin{equation}
{\frac{\partial }{\partial x^{1}}}(\frac{\mathcal{H}}{T})={\frac{\partial }{
\partial x^{2}}}(\frac{\mathcal{G}}{T}),\qquad {\frac{\partial }{\partial
x^{1}}}(\frac{\mathcal{G}}{T})=-{\frac{\partial }{\partial x^{2}}}(\frac{
\mathcal{H}}{T}).  \label{difconstr2}
\end{equation}
The scalars $f$ and $X^{I}$ are constrained via 
\begin{eqnarray}
{\frac{\partial }{\partial x^{1}}}(\frac{X_{I}}{f^{2}}) &=&\sqrt{2}i\chi T
\mathcal{H}V_{I},  \notag \\
\frac{\partial }{\partial x^{2}}(\frac{X_{I}}{f^{2}}) &=&\sqrt{2}i\chi T
\mathcal{G}V_{I}.
\end{eqnarray}
and $T$ satisfies 
\begin{equation}
\Box \log T=18\frac{\chi ^{2}}{f^{2}}(X^{I}X^{J}-{\frac{1}{2}}
Q^{IJ})V_{I}V_{J}T^{2}.
\end{equation}
Finally, $\Omega $ is constrained by 
\begin{eqnarray}
d\Omega &=&-{\frac{i}{\sqrt{2}}}(T{\frac{\partial \mathcal{G}}{\partial x^{2}
}}+\mathcal{H}{\frac{\partial T}{\partial x^{1}}})\left( dx^{1}\wedge dx^{2}-
\frac{1}{T^{2}}d\phi \wedge d\psi \right)  \notag \\
&-&\frac{3\chi }{f^{4}}V_{I}X^{I}\left( T^{2}dx^{1}\wedge dx^{2}+d\phi
\wedge d\psi \right) ,  \label{domegsol3}
\end{eqnarray}
and the gauge field strengths are then given by 
\begin{equation}
F^{I}=d\left( f^{2}X^{I}(dt+\Omega )\right) +6\chi \frac{T^{2}}{f^{2}}
V_{J}(X^{I}X^{J}-{\frac{1}{2}}Q^{IJ})dx^{1}\wedge dx^{2}.
\end{equation}

\section{Simplification of the Solutions}

The solutions of type $(2)$, $(3)$, $(4)$ are given in the most explicit
possible form. Hence, we shall concentrate on the solutions of type $(1)$, $
(5)$, $(6)$.

\subsection{Simplification of type (1) Solutions}

We find it convenient to simply the expression for the metric and gauge
field strengths of these solutions by changing co-ordinates from $(t,\tau
,\eta ,u,v)$ to $(t^{\prime },\phi ,w,u,v)$. There are two cases,
corresponding to $\theta \neq 0$ and $\theta =0.$

\subsubsection{Solutions with $\protect\theta \neq 0$}

If $\theta \neq 0$, it is convenient to make the co-ordinate transformation

\begin{eqnarray}
t &=&t^{\prime }-{\frac{w}{2c}}(H^{2}-c^{2}v^{2}f^{-6})\cos Y-{\frac{
wv\theta }{2c},}  \notag \\
\eta &=&v\sin Yw,  \notag \\
\tau &=&\phi +{\frac{w}{\theta }}(H^{2}-c^{2}v^{2}f^{-6})+vw\cos Y.
\end{eqnarray}

In these new co-ordinates, the solution is specified by: 
\begin{eqnarray}
ds_{\mathbf{B}}^{2} &=&H^{2}(d\phi +(v\cos Y+\theta
^{-1}(H^{2}-c^{2}v^{2}f^{-6}))dw)^{2}  \notag \\
&+&H^{-2}dv^{2}+H^{2}v^{2}\sin ^{2}Y(dw^{2}+du^{2})  \notag \\
dt+\Omega  &=&dt^{\prime }-{\frac{1}{2cv}}(H^{2}+c^{2}v^{2}f^{-6})d\phi  
\notag \\
&-&\big({\frac{1}{2c\theta v}}(H^{4}-c^{4}v^{4}f^{-12})+{\frac{1}{c}}
(H^{2}\cos Y+{\frac{\theta v}{2}})\big)dw  \notag \\
F^{I} &=&d\big(f^{2}X^{I}(dt^{\prime }-{\frac{1}{2cv}}
(H^{2}-c^{2}v^{2}f^{-6})d\phi   \notag \\
&-&({\frac{1}{2cv\theta }}(H^{2}-c^{2}v^{2}f^{-6})^{2}+{\frac{H^{2}}{c}}\cos
Y+{\frac{\theta v}{2c}}+cv^{2}f^{-6})dw)\big)  \notag \\
X_{I} &=&f^{2}\left( {\frac{q_{I}}{v}+\frac{\chi }{c}}\left( \frac{c^{2}v}{
f^{6}\theta }-{\frac{H^{2}}{\theta v}}-1\right) V_{I}\right) .
\end{eqnarray}

for constants $q_I$, where $Y, H$ are functions of $u, v$ ($\sin Y \neq 0$)
satisfying the constraints ({\ref{Hcons2}}) and ({\ref{Ycons2}}). Note that $
{\frac{\partial }{\partial t^{\prime}}}$, ${\frac{\partial }{\partial \phi}}$
and ${\frac{\partial }{\partial w}}$ are commuting Killing vectors which are
also symmetries of the full solution.

\subsubsection{Solutions with $\protect\theta=0$}

In the special case when $\theta =0$, note that 
\begin{equation}
d({\frac{c}{\chi }}vf^{-2}X_{I}+vV_{I})=V_{I}(\cos Ydv+H^{2}v\sin ^{2}Ydu)
\end{equation}
as not all of the $V_{I}$ vanish, fix some ${\tilde{I}}$ with $V_{\tilde{I}
}\neq 0$, so that 
\begin{equation}
d({\frac{c}{\chi }}vf^{-2}{\frac{X_{\tilde{I}}}{V_{\tilde{I}}}}+v)=\cos
Ydv+H^{2}v\sin ^{2}Ydu.
\end{equation}
It is also convenient to define 
\begin{equation}
X=vf^{-2}{\frac{X_{\tilde{I}}}{V_{\tilde{I}}}.}
\end{equation}
The co-ordinate transformation is then given by

\begin{eqnarray}
t &=&t^{\prime }-{\frac{w}{2c}}(H^{2}-c^{2}v^{2}f^{-6})\cos Y,  \notag \\
\eta  &=&v\sin Yw,  \notag \\
\tau  &=&\phi -w(v+{\frac{c}{\chi }}X)+vw\cos Y.
\end{eqnarray}
In these new co-ordinates, the solution is specified by 
\begin{eqnarray}
ds_{\mathbf{B}}^{2} &=&H^{2}(d\phi +(v(\cos Y-1)-{\frac{c}{\chi }}X)dw)^{2} 
\notag \\
&+&H^{-2}dv^{2}+H^{2}v^{2}\sin ^{2}Y(dw^{2}+du^{2}),  \notag \\
dt+\Omega  &=&dt^{\prime }-{\frac{1}{2cv}}(H^{2}+c^{2}v^{2}f^{-6})d\phi  
\notag \\
&+&\big({\frac{1}{2cv}}(H^{2}+c^{2}v^{2}f^{-6})(v+{\frac{c}{\chi }}X)-{\frac{
H^{2}}{c}}\cos Y\big)dw,  \notag \\
F^{I} &=&d\big(f^{2}X^{I}(dt^{\prime }-{\frac{1}{2cv}}
(H^{2}-c^{2}v^{2}f^{-6})d\phi   \notag \\
&+&({\frac{1}{2cv}}(H^{2}-c^{2}v^{2}f^{-6})(v+{\frac{c}{\chi }}X)-{\frac{
H^{2}}{c}}\cos Y-cv^{2}f^{-6})dw)\big),  \notag \\
vf^{-2}X_{I} &=&XV_{I}+q_{I},
\end{eqnarray}
for constants $c$, $q_{I}$ ($q_{\tilde{I}}=0$), where $Y,H$ are functions of 
$u,v$ ($\sin Y\neq 0$) satisfying the constraints ({\ref{Hcons2}}) and ({\ref
{Ycons2}}) with $\theta =0$.

Again, note that ${\frac{\partial }{\partial t^{\prime}}}$, ${\frac{\partial 
}{\partial \phi}}$ and ${\frac{\partial }{\partial w}}$ are commuting
Killing vectors which are also symmetries of the full solution.

\subsection{Simplification of type $(5)$ Solutions}

To simplify the solutions further, define 
\begin{equation}
Q=-{\frac{i}{\sqrt{2}}}\varrho (T^{-1}{\frac{\partial \mathcal{G}}{\partial
x^{2}}}+\mathcal{H}T^{-2}{\frac{\partial T}{\partial x^{1}}})+3\chi
u^{-2}V_{I}X^{I}
\end{equation}
so that 
\begin{equation}
\Omega =-{\frac{e^{-\sqrt{2}\varrho ^{2}\psi }}{\sqrt{2}\varrho ^{2}}}\big[
iT\varrho ^{3}(-\mathcal{G}dx^{1}+\mathcal{H}dx^{2})+Q(d\phi +\beta )\big]
\end{equation}
and noting that 
\begin{equation}
dQ=\sqrt{2}i\varrho ^{5}(T\mathcal{H}dx^{1}+T\mathcal{G}dx^{2})
\end{equation}
one finds 
\begin{equation}
d\big[Q^{2}-\varrho ^{6}({\frac{2u^{-6}}{\varrho ^{2}}}+\mathcal{G}^{2}+
\mathcal{H}^{2})\big]=0
\end{equation}
and hence 
\begin{equation}
Q^{2}=\xi +\varrho ^{6}({\frac{2u^{-6}}{\varrho ^{2}}}+\mathcal{G}^{2}+
\mathcal{H}^{2})
\end{equation}
for constant $\xi $. Also note that 
\begin{equation}
u^{-2}X_{I}={\frac{\chi }{\varrho ^{4}}}QV_{I}+q_{I}\text{ .}
\end{equation}
There are then a number of subclasses of solutions, according as to whether $
\mathcal{H}^{2}+\mathcal{G}^{2}\neq 0$ or $\mathcal{H}=\mathcal{G}=0$.

\subsubsection{Solutions with $\mathcal{H}^2 + \mathcal{G}^2 \neq 0$}

Suppose we consider a neighbourhood in which $\mathcal{H}^{2}+\mathcal{G}
^{2}\neq 0$. Note that as $T^{-1}(\mathcal{H}+i\mathcal{G})$ is a
holomorphic function of $x^{1}+ix^{2}$, it follows that ${\frac{T}{\mathcal{H
}+i\mathcal{G}}}$ is also a holomorphic function of $x^{1}+ix^{2}$, and so 
\begin{equation}
{\frac{\partial }{\partial x^{1}}}\left( {\frac{T\mathcal{H}}{\mathcal{H}
^{2}+\mathcal{G}^{2}}}\right) =-{\frac{\partial }{\partial x^{2}}}\left( {
\frac{T\mathcal{G}}{\mathcal{H}^{2}+\mathcal{G}^{2}}}\right) ,\qquad {\frac{
\partial }{\partial x^{1}}}\left( {\frac{T\mathcal{G}}{\mathcal{H}^{2}+
\mathcal{G}^{2}}}\right) ={\frac{\partial }{\partial x^{2}}}\left( {\frac{T
\mathcal{H}}{\mathcal{H}^{2}+\mathcal{G}^{2}}}\right)
\end{equation}
or equivalently

\begin{eqnarray}
d\left( {\frac{T\mathcal{G}}{\mathcal{H}^{2}+\mathcal{G}^{2}}}dx^{1}-{\frac{T
\mathcal{H}}{\mathcal{H}^{2}+\mathcal{G}^{2}}}dx^{2}\right)  &=&0,  \notag \\
d\left( {\frac{T\mathcal{H}}{\mathcal{H}^{2}+\mathcal{G}^{2}}}dx^{1}+{\frac{T
\mathcal{G}}{\mathcal{H}^{2}+\mathcal{G}^{2}}}dx^{2}\right)  &=&0,
\end{eqnarray}
hence one obtains (locally) real functions $z=z(x^{1},x^{2})$, $
y=y(x^{1},x^{2})$ such that 
\begin{eqnarray}
{\frac{T\mathcal{G}}{\mathcal{H}^{2}+\mathcal{G}^{2}}}dx^{1}-{\frac{T
\mathcal{H}}{\mathcal{H}^{2}+\mathcal{G}^{2}}}dx^{2} &=&idz,  \notag \\
{\frac{T\mathcal{H}}{\mathcal{H}^{2}+\mathcal{G}^{2}}}dx^{1}+{\frac{T
\mathcal{G}}{\mathcal{H}^{2}+\mathcal{G}^{2}}}dx^{2} &=&idy,
\end{eqnarray}
with

\begin{eqnarray}
{\frac{\partial }{\partial z}} &=&{\frac{i}{T}}\left( \mathcal{G}{\frac{
\partial }{\partial x^{1}}}-\mathcal{H}{\frac{\partial }{\partial x^{2}}}
\right) ,  \notag \\
{\frac{\partial }{\partial y}} &=&{\frac{i}{T}}\left( \mathcal{H}{\frac{
\partial }{\partial x^{1}}}+\mathcal{G}{\frac{\partial }{\partial x^{2}}}
\right) .
\end{eqnarray}
Next, note that one can solve for the 1-form $\beta $ to find (up to a total
derivative which can be neglected)

\begin{equation}
\beta =-{\frac{i}{\varrho ^{3}}}{\frac{QT}{\mathcal{H}^{2}+\mathcal{G}^{2}}}
(-\mathcal{G}dx^{1}+\mathcal{H}dx^{2})=-{\frac{Q}{\varrho ^{3}}}dz
\end{equation}
and the expression for $\Omega $ can be further simplified to

\begin{equation}
\Omega =-{\frac{1}{\sqrt{2}\varrho ^{2}}}e^{-\sqrt{2}\varrho ^{2}\psi
}\left( Qd\phi -{\frac{1}{\varrho ^{3}}}(\xi +2\varrho ^{2}u^{-6})dz\right) .
\end{equation}
It is then straightforward to see that ${\frac{\partial }{\partial z}}$ is
an additional Killing vector, which is also a symmetry of the full solution: 
$Q,X^{I},u$ are functions only of $y$, and the metric on the K\"{a}hler base
space simplifies to

\begin{equation}
ds_{\mathbf{B}}^{2}=e^{\sqrt{2}\varrho ^{2}\psi }\big((d\phi -{\frac{Q}{
\varrho ^{3}}}dz)^{2}+d\psi ^{2}-(\mathcal{H}^{2}+\mathcal{G}
^{2})(dz^{2}+dy^{2})\big).
\end{equation}
Finally, it is most useful to change co-ordinates from $(t,\psi ,\phi ,z,y)$
to $(t,\psi ,\phi ,z,Q)$; where

\begin{equation}
dQ=-\sqrt{2}\varrho ^{5}(\mathcal{H}^{2}+\mathcal{G}^{2})dy.
\end{equation}
In these new co-ordinates, the solution is given by

\begin{eqnarray}
ds_{\mathbf{B}}^{2} &=&e^{\sqrt{2}\varrho \psi }\bigg[(d\phi -{\frac{Q}{
\varrho ^{3}}}dz)^{2}+d\psi ^{2}+(2u^{-6}\varrho ^{-2}-\varrho
^{-6}(Q^{2}-\xi ))dz^{2}  \notag \\
&+&{\frac{1}{2(2u^{-6}\varrho ^{8}-\varrho ^{4}(Q^{2}-\xi ))}}dQ^{2}\bigg], 
\notag \\
u^{-2}X_{I} &=&{\frac{\chi }{\varrho ^{4}}}QV_{I}+q_{I},  \notag \\
\Omega  &=&-{\frac{1}{\sqrt{2}\varrho ^{2}}}e^{-\sqrt{2}\varrho ^{2}\psi
}(Qd\phi -{\frac{1}{\varrho ^{3}}}(\xi +2\varrho ^{2}u^{-6})dz),  \notag \\
F^{I} &=&d\bigg(u^{2}e^{\sqrt{2}\varrho ^{2}\psi }X^{I}(dt+\Omega )\bigg)+3
\sqrt{2}\chi \varrho ^{-5}u^{-2}V_{I}(X^{I}X^{J}-{\frac{1}{2}}
Q^{IJ})dz\wedge dQ,  \notag \\
f &=&e^{{\frac{\varrho ^{2}}{\sqrt{2}}}\psi }u.
\end{eqnarray}
for constants $\varrho \neq 0$, $q_{I}$, $\xi $. Note that ${\frac{\partial 
}{\partial t}},{\frac{\partial }{\partial \phi }},{\frac{\partial }{\partial
z}}$ are commuting Killing vectors which are symmetries of the full solution.

\subsubsection{Solutions with $\mathcal{H}=\mathcal{G}=0$}

If $\mathcal{H}=\mathcal{G}=0$ then the scalars $X^{I}$ are constant, as is $
u$. Without loss of generality, set $u=1$. With these constraints, the
function $T$ which is introduced \cite{halfgs} must satisfy 
\begin{equation}
T^{-2}\Box \log T=\Lambda   \label{lapT}
\end{equation}
where 
\begin{equation*}
\Lambda =-2\varrho ^{4}+18\chi ^{2}(X^{I}X^{J}-{\frac{1}{2}}Q^{IJ})V_{I}V_{J}
\end{equation*}
is constant. Let $M$ be a 2-manifold equipped with metric 
\begin{equation}
ds^{2}(M)=T^{2}((dx^{1})^{2}+(dx^{2})^{2}).
\end{equation}
Then ({\ref{lapT}}) implies that $M$ has Ricci scalar ${}^{(M)}R=-2\Lambda $
, so is isometric to $\mathbb{S}^{2}$, $\mathbb{R}^{2}$ or $\mathbb{H}^{2}$
according as $\Lambda <0$, $\Lambda =0$ or $\Lambda >0$ respectively. Note
also that the 1-form $\beta $ must satisfy 
\begin{equation}
d\beta =\sqrt{2}\varrho ^{2}\mathrm{dvol}\ (M).
\end{equation}
Hence, to summarize, these solutions have constant $X^{I}$, and 
\begin{eqnarray}
ds_{\mathbf{B}}^{2} &=&e^{\sqrt{2}\varrho ^{2}\psi }((d\phi +\beta
)^{2}+d\psi ^{2}+ds^{2}(M)),\qquad d\beta =\sqrt{2}\varrho ^{2}\mathrm{dvol}
\ (M),  \notag \\
{}^{(M)}R &=&4\varrho ^{4}-36\chi ^{2}(X^{I}X^{J}-{\frac{1}{2}}
Q^{IJ})V_{I}V_{J},  \notag \\
\Omega  &=&-3\chi V_{I}X^{I}{\frac{e^{-\sqrt{2}\varrho ^{2}\psi }}{\sqrt{2}
\varrho ^{2}}}(d\phi +\beta ),  \notag \\
F^{I} &=&d\bigg(e^{\sqrt{2}\varrho ^{2}\psi }X^{I}(dt+\Omega )\bigg)+6\chi
(X^{I}X^{J}-{\frac{1}{2}}Q^{IJ})V_{J}\mathrm{dvol}\ (M),  \notag \\
f &=&e^{{\frac{\varrho ^{2}}{\sqrt{2}}}\psi }.
\end{eqnarray}
where $\varrho $ is a non-zero constant.

Note that ${\frac{\partial }{\partial t}}$ and ${\frac{\partial }{\partial
\phi}}$ are commuting Killing vectors which are symmetries of the full
solution. A Killing vector of $M$ can also be obtained which commutes with
both ${\frac{\partial }{\partial t}}$ and ${\frac{\partial }{\partial \phi}}$
and is also a symmetry of the full solution. So again, there are three
commuting Killing vectors which are symmetries of the full solution.

\subsection{Simplification of type $(6)$ Solutions}

Again, for these solutions, there are two sub-classes, according as $
\mathcal{G}^2+\mathcal{H}^2 \neq 0$, or $\mathcal{G}=\mathcal{H}=0$.

\subsubsection{Solutions with $\mathcal{H}^2 + \mathcal{G}^2 \neq 0$}

In order to simplify the solutions of type $(6)$, note that the functions $
\mathcal{G}, \mathcal{H}, T$ satisfy the same constraints ({\ref{difconstr1}}
) and ({\ref{difconstr2}}) as the type $(5)$ solutions, and hence we again
introduce co-ordinates $z$, $y$ such that

\begin{eqnarray}
{\frac{T\mathcal{G}}{\mathcal{H}^{2}+\mathcal{G}^{2}}}dx^{1}-{\frac{T
\mathcal{H}}{\mathcal{H}^{2}+\mathcal{G}^{2}}}dx^{2} &=&idz,  \notag \\
{\frac{T\mathcal{H}}{\mathcal{H}^{2}+\mathcal{G}^{2}}}dx^{1}+{\frac{T
\mathcal{G}}{\mathcal{H}^{2}+\mathcal{G}^{2}}}dx^{2} &=&idy,
\end{eqnarray}
with

\begin{eqnarray}
{\frac{\partial }{\partial z}} &=&{\frac{i}{T}}\big(\mathcal{G}{\frac{
\partial }{\partial x^{1}}}-\mathcal{H}{\frac{\partial }{\partial x^{2}}}
\big),  \notag \\
{\frac{\partial }{\partial y}} &=&{\frac{i}{T}}\big(\mathcal{H}{\frac{
\partial }{\partial x^{1}}}+\mathcal{G}{\frac{\partial }{\partial x^{2}}}
\big).
\end{eqnarray}
It is also convenient to define the scalar $Q$ by 
\begin{equation}
Q^{2}=-(\mathcal{G}^{2}+\mathcal{H}^{2}).
\end{equation}
Then the constraints ({\ref{difconstr1}}) and ({\ref{difconstr2}}) imply
that $Q$, $f$ and $X^{I}$ are functions only of $y$, with

\begin{equation}
{\frac{d}{dy}}(f^{-2}X_{I})=\sqrt{2}\chi Q^{2}V_{I}.  \label{auxsc1}
\end{equation}
Next, note that

\begin{eqnarray}
d\Omega &=&\bigg({\frac{1}{\sqrt{2}}}{\frac{d\log Q}{dy}}-3\chi
f^{-4}V_{I}X^{I}\bigg)d\phi \wedge d\psi  \notag \\
&+&\bigg(-{\frac{1}{\sqrt{2}}}{\frac{d\log Q}{dy}}-3\chi f^{-4}V_{I}X^{I}
\bigg)T^{2}dx^{1}\wedge dx^{2}.
\end{eqnarray}
\ \ The integrability condition of this constraint is given by 
\begin{equation}
{\frac{1}{\sqrt{2}}}{\frac{d\log Q}{dy}}-3\chi f^{-4}V_{I}X^{I}=\xi
\label{aux2}
\end{equation}
for constant $\xi $.

It is then straightforward to show that 
\begin{equation}
d\Omega =\xi d\phi \wedge d\psi +d\big({\frac{1}{\sqrt{2}}}({\frac{1}{2}}
Q^{2}+f^{-6})dz\big)
\end{equation}
and also 
\begin{equation}
F^{I}=d\bigg(f^{2}X^{I}(dt+\Omega )-\sqrt{2}f^{-4}X^{I}dz\bigg).
\end{equation}

Finally, consider the constraints ({\ref{auxsc1}}) and ({\ref{aux2}}). As
not all $V_{I}$ vanish, choose ${\tilde{I}}$ such that $V_{\tilde{I}}\neq 0$
, then ({\ref{auxsc1}}) implies 
\begin{equation}
Q^{2}={\frac{1}{\sqrt{2}\chi }}{\frac{d}{dy}}\bigg(f^{-2}{\frac{X_{\tilde{I}}
}{V_{\tilde{I}}}}\bigg).
\end{equation}
It is then convenient to define 
\begin{equation}
X=f^{-2}{\frac{X_{\tilde{I}}}{V_{\tilde{I}}}}\ .
\end{equation}

Combining all of the above constraints, one finds that the solution is
specified by 
\begin{eqnarray}
f^{-2}X_{I} &=&XV_{I}+q_{I},\quad \mathrm{where}\ X\ \mathrm{satisfies}\quad 
{\frac{1}{4\chi }}{\frac{dX}{dy}}-{\frac{1}{2}}f^{-6}={\frac{\xi }{\sqrt{2}
\chi }}X,  \notag \\
ds_{\mathbf{B}}^{2} &=&d\phi ^{2}+d\psi ^{2}+2\sqrt{2}({\frac{1}{2}}f^{-6}+{
\frac{\xi }{\sqrt{2}\chi }}X)(dy^{2}+dz^{2}),  \notag \\
d\Omega  &=&\xi d\phi \wedge d\psi +d\bigg((\sqrt{2}f^{-6}+{\frac{\xi }{
\sqrt{2}\chi }}X)dz\bigg),  \notag \\
F^{I} &=&d\bigg(f^{2}X^{I}(dt+\Omega )-\sqrt{2}f^{-4}X^{I}dz\bigg).
\end{eqnarray}
for constants $q_{I}$ ($q_{\tilde{I}}=0$), $\xi $.

Note that ${\frac{\partial }{\partial t}}, {\frac{\partial }{\partial \phi}}
, {\frac{\partial }{\partial z}}$ are commuting Killing vectors which are
symmetries of the full solution.

\subsubsection{Solutions with $\mathcal{H}=\mathcal{G}=0$}

For these solutions, the scalars $X^{I}$ and $f$ are constant; without loss
of generality set 
\begin{equation}
f=1.
\end{equation}
Then 
\begin{equation}
ds_{B}^{2}=d\phi ^{2}+d\psi ^{2}+ds^{2}(M)
\end{equation}
where $M$ is a 2-manifold which is either $\mathbb{S}^{2}$, $\mathbb{R}^{2}$
or $\mathbb{H}^{2}$ according as the Ricci scalar 
\begin{equation*}
{}^{(M)}R=-36\chi ^{2}(X^{I}X^{J}-{\frac{1}{2}}Q^{IJ})V_{I}V_{J}
\end{equation*}
is positive, zero, or negative. In addition, one has 
\begin{eqnarray}
d\Omega &=&-3\chi V_{I}X^{I}(\mathrm{dvol\ }(M)+d\phi \wedge d\psi )  \notag
\\
F^{I} &=&-3\chi X^{I}X^{J}V_{J}d\phi \wedge d\psi +3\chi
(X^{I}X^{J}-Q^{IJ})V_{J}\mathrm{dvol\ }(M).
\end{eqnarray}
It is clear that this solution also admits three Killing vectors which are
symmetries of the full solution.

\bigskip

\section{Summary}

In summary we have revisited the classification of $1/2$ supersymmetric
solutions of the theory of $N=2,D=5$ supergravity which have at least one
time-like Killing spinor. Three of the six classes of these solutions were
given in their most explicit form in \cite{halfgs}. Our purpose was to
recast the remaining three solutions in a form which enabled us to extract
some hidden isometries of the metric solutions. We found coordinate
transformations that simplified these three classes of solutions and allowed
the explicit construction of their Killing vectors. It is of interest to
investigate whether there are any regular asymptotically $AdS_{5}$ black
ring solutions. Supersymmetric rings exist in the ungauged theory 
\cite{harveyelv:04, harveyelv:04b, ggbr1:04, ggbr2:04} and it is known that
supersymmetry is fully restored at the ring horizon. 
Furthermore, supersymmetric regular asymptotically $AdS_5$ black holes
undergo supersymmetry enhancement from $1/4$ to $1/2$ supersymmetry
in their near horizon limits.
So, if $AdS_{5}$ black
rings exist in the gauged theory, one might also expect that
supersymmetry is enhanced from $1/4$ to $1/2$ at the horizon.

Recent work 
\cite{har} has shown that there are no black rings which have
horizons with a $\left[ U(1)\right]
^{2}$ symmetry. In our work, we have shown that 
the $1/2$ supersymmetric solutions in the ``timelike" class
possess, in addition to the timelike Killing vector generated by
the timelike Killing spinor, two further commuting Killing vectors. However,
 these two additional Killing vectors are {\it not} necessarily spacelike.
 Hence, in order to determine if there exists
 a $1/2$ supersymmetric solution corresponding to
the near-horizon geometry of a black ring, it will be necessary to
construct a more detailed analysis of half-supersymmetric near horizon
geometries. The simplifications to the half-supersymmetric solutions
presented in this work will be useful in constructing such a classification.

\vskip1cm

{\flushleft{\textbf{Acknowledgments}}}

{\flushleft{The work of W. A. Sabra was supported in part by the National Science
Foundation under grant number PHY-0703017.}}

\bigskip

\end{document}